\documentclass[graybox]{svmult}

\usepackage{mathptmx}       % selects Times Roman as basic font
\usepackage{helvet}         % selects Helvetica as sans-serif font
\usepackage{courier}        % selects Courier as typewriter font
\usepackage{type1cm}        % activate if the above 3 fonts are
\usepackage{makeidx}         % allows index generation
\usepackage{graphicx}        % standard LaTeX graphics tool
\usepackage{multicol}        % used for the two-column index
\usepackage[bottom]{footmisc}% places footnotes at page bottom

\usepackage{algorithm}
\usepackage{algorithmic}

\usepackage{float}

\usepackage{longtable}

\makeindex             % used for the subject index
\begin{document}
\title{Continuous Features Discretizaion for Anomaly Intrusion Detectors Generation}
\author{Amira Sayed A.Aziz, Ahmad Taher Azar, Aboul ella Hassanien, Sanaa Al-Ola Hanafy}

\institute{Amira Sayed A.Aziz \at Scientific Research Group  in Egypt (SRGE), Cairo, Egypt, \email{amira.abdelaziz@scienceegypt.net}
\and Ahmad Taher Azar \at PhD, IEEE Senior Member, Faculty of Computers and Information, Benha University, Egypt. \email{ahmad.azar@fci.bu.edu.eg}
\and Aboul ella Hassanien \at Scientific Research Group  in Egypt (SRGE), Cairo, Egypt, \email{Aboulella@scienceegypt.net}
\and Sanaa Al-Ola Hanafy \at Cairo University, Faculty of Computers and Information, Cairo, Egypt}
\maketitle

\abstract {Network security is a growing issue, with the evolution of computer systems and expansion of attacks. Biological systems have been inspiring scientists and designs for new adaptive solutions, such as genetic algorithms. In this paper, we present an approach that uses the genetic algorithm to generate anomaly network intrusion detectors. In this paper, an algorithm propose use a discretization method for the continuous features selected for the intrusion detection, to create some homogeneity between values, which have different data types. Then,the intrusion detection system is tested against the NSL-KDD data set using different distance methods. A comparison is held amongst the results, and it is shown by the end that this proposed approach has good results, and recommendations is given for future experiments.}

\section{Introduction}
With the evolution of computer networks during the past few years, security is a crucial issue and a basic demand for computer systems. Attacks are expanding and evolving as well, making it important to find and work on new and advanced solutions for network security. Intrusion Detection Systems (IDS) have been around us for a some time, as an essential mechanism to protect computer systems, where they identify malicious activities that occur in that protected system.
Genetic Algorithms (GA) are a group of computational models inspired by natural selection. This solution works on a group of chromosomes-like data structure (a population) where they reproduce and the new generation replaces the old one if it's more fitting in the environment. These new generations are generated using selection and recombination functions such as crossover and mutation \cite{11}.
The GAs were first seen as optimization solutions, but now they are applied in a variety of systems, including the IDSs. The GA then is used as a machine learning technique to generate artificial intelligence detection rules. The rules are usually in the if-then forms, where the conditions are values that represent normal samples or values to indicate an intrusion is in the act \cite{11}, \cite{3}. For a Network-based IDS (NIDS), usually the network traffic is used to build a model and detect anomalous network activities. Many features can be extracted and used in a GA to generate the rules, and these features may be of different data types, and may have a wide range of values. So, in our paper we propose using a discretization algorithm with continuous features to create homogeneity amongst features. \\
---add more in introduction about discretization.\\
The rest of this paper is organized as follows: Section 1.2 gives a background of the different algorithms used in our approach. Section 1.3 gives a review on some of the previous work done in the area. Section 1.4 describes the proposed approach, Section 1.5 gives an overview of the experiment and results, and we list our conclusions and future work in Section 1.6.

\section{Background}
\subsection{Anomaly Intrusion Detection}
Intrusion Detection Systems (IDSs) are security tools used to detect anomalous or malicious activity from inside and outside intruders. An IDS can be host-based or network-based, which is the concern in this paper. They are classified by many axes, one of them is the detection methodology that classifies them to signature-based and anomaly-based IDS. The former detects attacks by comparing the data to patterns stored in a signature database of known attacks. The later detects anomalies by defining a model of normal behaviour of the monitored system, then considers any behavior lying outside the model as anomalous or suspicious activity. Signature-based IDS can detect well-known attacks with high accuracy but fails to detect or find unknown attacks. Anomaly-based IDS has the ability to detect new or unknown attacks but usually has high false positives rate (normal activities detected as anomalous). There are three types of anomaly detection techniques: statistical-based, knowledge-based, and machine learning-based, as shown in the figure below. We can measure an IDS performance by two key aspects: the detection process efficiency and the involved cost of the operation \cite{1}.

\subsection{Genetic Algorithms}
Genetic Algorithm (GA) is an evolutionary computational technique that is used as a search algorithm, based on the concepts of natural selection and genetics. There are 3 meanings of search:
1- Search for stored data: where the problem is to retrieve some information stored in a computer memory efficiency.
2- Search for paths to goals: where one needs to find the best paths from an initial state to a goal.
3- Search for solutions: where one needs to find a solution or group of solutions in a large space of candidates.
GA works on a population of individuals, where each individual is called a chromosome and is composed of a string of values called genes. The population goes through a process to find a solution or group of high quality solutions. The quality of an individual is measured by a fitness function that is dependant on the environment and application. The process starts with an initial population, that goes through transformation for a number of generations. During each generation, three major operations are applied sequentially to each individual: selection, crossover, and mutation. This is shown in Figure 1.1 \cite{2}\cite{3}.

\begin{figure}[h]
\begin{center}
{\includegraphics[width=2.9in,height=3.3in]{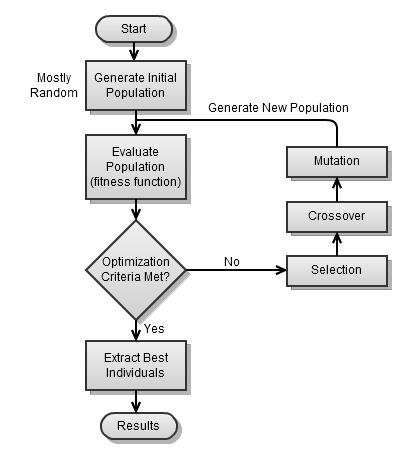}}
\caption{Genetic Algorithm Process}
\end{center}
\end{figure}

\subsection{Negative Selection Approach}
Artificial Immune Systems (AIS) are inspired by the nature's Human Immune System (HIS), which is an adaptive, tolerant, self-protecting, and dynamic defence system. AIS is a set of algorithms that mimic the different functionalities of the HIS, and they can perform a range of tasks. The major algorithms are: negative selection, clonal selection, and immunity networks. The Negative Selection Approach (NSA) is based on the concept of self-nonself discrimination, by first creating a profile of the self (normal) behaviour and components. Then use this profile to rule out any behaviour that doesn't match with that profile. The training phase goes on the self samples. Then, the detectors are exposed to different samples, and if a detector matches a self as nonself then it's discarded. The final group of detectors (mature detectors) are released to start the detection process \cite{4}.

\subsection{Equal-Width Binning Algorithm}
There are many discretization algorithms for continuous features to discretize them for algorithmic purposes. These discretization algorithms are very important for machine learning, as tis is required by some algorithms. But more importantly, the discretization increases the speed of induction algorithms. The discretization algorithms are classified in many ways --- as shown in Table 1 below \cite{5} \cite{14} \cite{15}.

\begin{table}[h]
\caption{Discretization Algorithms Classification}
\begin{center}
\begin{tabular}{| p{4cm} | p{7cm} |}
\hline Local vs. Global & Local methods apply partitions on localized regions of the instance space. Global methods works on the entire instance space, that every feature is partitioned into number of regions independent of other attributes. \\ 
\hline Supervised vs. Unsupervised & Supervised methods make use of the class labels associated with the instances in the process, while unsupervised methods perform discretization regardless of class label. \\ 
\hline Static vs. Dynamic & Static methods determine the number of bins for each feature independent of other features after performing one discretization pass of the data (performed before the classification). Dynamic methods determine the number of bins for all features simultaneously by a search through the data space (performed while the classifier is built). \\ 
\hline Top-down(Splitting) vs. Bottom-up (Merging) & Splitting methods start with an empty group of cut points, and build up during the discretization process. While in merging, the algorithm starts with a list of cut points, then discards unneeded ones during discretization by merging intervals.\\
\hline Direct vs. Incremental & In direct methods, number of bins (intervals) is predefined either by user or using an algorithm. Incremental methods start with simple discretization that gets improved and refined until stopped by a conditon (meeting a certain criterion).\\
\hline 
\end{tabular} 
\end{center}
\end{table}

\section{Previous Work}
In this section, a number of papers which investigated the discretization algorithms, combining them with other methods for clustering and feature selection, are highlighted in Table 2.

%\begin{table}[h]
%\caption{Discretization Previous Work}
%\begin{center}
%\begin{tabular}{ p{2cm} p{0.2cm} p{4.3cm} p{0.2cm} p{4.3cm} }
\begin{longtable} { p{2cm} p{0.2cm} p{4.3cm} p{0.2cm} p{4.3cm} }
\hline Authors & & Technique & & Results \\ 
\hline J. Dougherty, R. Kohavi, M. Sahami \cite{16} & & Applied EWB, 1R, and Recursive Entropy Partitioning as preprocessing step before using C4.5 and Naive-Bayes classifiers on data. Data Set: 16 data sets from the UC Irvine (UCI) Repository. & & C4.5 performance. improved on 2 data sets using entropy discretization, but slightly decreased on some. At 95 \% confidence level, Naive-Bayes with entropy discretization is better than C4.5 on 5 data sets and worse on 2 (with average accuracy 83.97\% vs. 82.25\% for C4.5)\\
\hline Ellis J. Clarke, Bruce A. Barton \cite{17} & & They used Minimum Descriptive Length (MDL) to select number of intervals, and modified a version of the k2 method for one test, entropy based discretization for another test. Data Set: NGHS and DISC from two epidemiological studies. & & Dynamic Partitioning with MDL metric lead to more highly connected BBN than with only entropy partitioning. New proposed method lead to better representations of variable dependencies in both data sets. But generally, using entropy and MDL partitioning provided clarification and simplification in the BBN. \\
\hline Zhao Jun, Zhou Ying-hua \cite{18} & & A rough set based heuristic method, enhanced in two ways: (1) decision information is used in candidate cut computation (SACC), and (2) an estimation of cut selection probability is defined to measure cut significance (ABSP). Data Set: continuous UCI data sets. & & Their SACC was compared to an algorithm knwon as UACC, and ABSP was comapred to some typical rough set based discretization algorithm. SACC performs better with less number of cuts, and ABSP slightly improves predictive accuracies. \\
\hline Anika Gupta, Kishan Mehrotra, Chilukuri Mohan \cite{14} & & They applied clustering as a preprocessing step before the discretization process. For clustering they used: k-means with euclidean distance similarity metric, and Shared Nearest Neighbour (SNN). For discretization they used a generalized version of MDL with alpha=Beta=0.5 (ME-MDL). Data Set: 11 data sets from UCI repository. & & Comparing their algorithm results with ME-MDL results: in all data sets, when SNN or k-means clustering is used, the proposed algorithm gives better results. In heart data set, SNN clustering does better than ME-MDL. In other data sets, k-means does better than SNN.\\
\hline Daniela Joita \cite{19} & & A discretization algorithm based on the k-means clustering algorithm, and that avoids the O(n log n) time required for sorting. & & The algorithm  was proposed to be tested in the future.\\
\hline Shunling Chen, Liny Tang, Wei Jun Liu, Yonghong Li \cite{20} & &  An improved method of continuous attributes discretization by: (1) hierarchical clustering is applied to form initial division of the attribute, and (2) merging adjacent ranges based on entropy, taking into consideration not to affect level of consistency of the decision table. Data Set: data of their provincial educational committee project. & & Not listed, but mentioned to prove the validity of their algorithm.\\
\hline Artur J. Ferreira, Mario A.T. Figueirdo \cite{21} & & They used clustering with discretization for better results. They proposed an updated version of the well-know Linde-Buzp-Gray (LBG) algorithm: U-LBG1 (used a variable number of bits) and U-LBG2 (used a fixed number of bits). For clustering, they used the Relevance-Redundancy Feature Selection (RRFS) and Relevance Feature Selection (RFS) methods. Data Set: data sets from UCI, the five data sets of the NIP2003 FS challenge, and several micro-array gene expression data sets (no normalization was applied on any of the used data sets). & & The proposed approaches allocated a small number of bits per feature. RRFS performs better than RS for eliminating redundant features.\\
\hline 
\end{longtable}
%\end{tabular} 
%\end{center}
%\end{table}

Also very good reviews are given in \cite{15} and \cite{22}.

\section{Proposed Approach}
\subsection{Motivation}
In \cite{6} The algorithm was originally suggested with the application on real-valued features in the NSL-KDD data set. It was used with a variation parameter defining the upper and lower limits of the detectors values (conditions). It had very good results, but the real-valued features are not enough to detect all types of attacks, so the algorithm should expand to include features of different types. In \cite{7}, they applied the algorithm on the KDD data set, using a range of features to detect anomalies. The problem with using different features is that they have different data types ranges: binary, categorical, and continuous (real and integer). This may lead to problems while applying the algorithm. First of all, a wide range of values need to be covered in a way that can represent each region uniquely. Secondly, there should be some sort of homogeneity between features values to apply the GA. So, the use of some discretization algorithm for continuous features lead to the suggestion of our approach.

\subsection{Suggested Approach}
Continuous features discretization methods increase the speed of induction algorithms, beside being an algorithmic requirement in machine learning. Equal-width interval binning \cite{5} is the simplest method for data discretization, where the range of values is divided into k equally sized bins, as $k$ is a parameter supplied by the user as the required number of bins. The bin width is calculated as:
\begin{equation}
\delta = \frac{x_{max} - x_{min}}{k}
\end{equation}

and the bin boundaries are set as: $x_{min}$+i$\delta$, $i$=1,...,$k$-1.

The equal-width interval binning algorithm is global, unsupervised, and static discretization algorithm.

The suggested approach starts with binning the continuous features with a previously defined number of bins, Then, replace each feature value with its enclosing bin number. Finally, run the GA on the modified data set samples to generate the detectors (rules). Following the NSA concepts, this is applied on the normal samples through the training phase. The self samples are presented in the self space S. The process is shown in  Algorithm I.

\begin{algorithm}[h]
\caption{Proposed Algorithm} \label{Level set algorithm}
\begin{algorithmic}[1]
\STATE Run equal-width binning algorithm on continuous features.
\STATE	Initialize population by selecting random individuals from the space $S$.
\FOR{The specified number of generations}
\FOR{The size of the population}
\STATE{	Select two individuals (with uniform probability) as $parent_1$ and $parent_2$.}
\STATE	Apply crossover to produce a new individual ($child$).
	\STATE Apply mutation to child.
\STATE	Calculate the distance between $child$ and $parent_1$ as $d_1$, and the distance between $child$ and $parent_2$ as $d_2$.
\STATE Calculate the fitness of $child$, $parent_1$, and $parent_2$ as $f$, $f_1$, and $f_2$ respectively.
\IF {($d_1 < d_2$) and ($ f>f_1$)}
\STATE replace $parent_1$ with $child$
\ELSE
\IF {($d_2<=d_1$) and ($f>f_2$) }
\STATE Replace $parent_2$ with $child$.
\ENDIF
\ENDIF
\ENDFOR
\ENDFOR
\STATE Extract the best (highly-fitted) individuals as your final solution.
\end{algorithmic}
\end{algorithm}

The fitness - which was inspired from \cite{10} - is measured by calculating the matching percentage between an individual and the normal samples, as:
\begin{equation}
fitness(x) = \frac{a}{A}
\end{equation}
where $a$ is the number of samples matching the individual by 100\% , and $A$ is the total number of normal samples.
Three different distance methods were tested (one at a time), to find the best results. The distances measured between a child $X$ and a parent $Y$ using the following formulas:

\begin{itemize}

\item{The Euclidean distance as:
\begin{equation}
d(X,Y) = \sqrt{(x_1-y_1)^2 + (x_2-y_2)^2 \dots (x_n-y_n)^2}
\end{equation}}

\item{The Hamming distance, which defines the difference between 2 strings (usually binary) as the number of places in which the strings have different values\cite{12}. So it's calculated as (where n is number of features):
\begin{equation}
d(X,Y) = \sum\limits_{i=0}^{n} \left | (x_i-y_i) \right |
\end{equation} }

\item{The Minkowski Distance, which is similar to the Euclidean distance but uses the p-norm dimension as the power value instead. So, the formula goes as:
\begin{equation}
d(X,Y) = (\sum_{i=0}^{n} (\left | x_{i} - y_{i} \right |^{p}))^{1/p}
\end{equation}
}
The euclidean distance uses the same equation with $p$=2. If - for example - $p$=1, then it's the Manhattan distance. In the Minkowski distance case, $p$ can be any value larger than 0 and up to infinity. It can be have real value between 0 and 1. If we are interested in finding the difference between objects, then we should aim for high $p$ values. If we are interested in finding the how much the objects are similar, then we should go for low $p$ values \cite{13}. In our experiment, a small value of 0.5 was used, and a big values of 18 was used to compare results.
\end{itemize} 

\section{Experiment}
\subsection{Data Set}
The NSL-KDD IDS data set \cite{8} was proposed in \cite{9} to solve some issues in the widely use KDD Cup 99 data set. These issues affect the performance of the systems that use the KDD data set and results in very poor evaluation of them. The resulted data set is having a reasonable size and is unbiased, and it's affordable to use in the experiments without having to select a small portion of the data. The data sets used in our experiment are:
\begin{itemize}
\item KDDTrain+\_20Percent normal samples for training and generating the detectors.
\item KDDTrain+ and KDDTest+ for testing, where the difference between them is that the Test set include additional unknown attacks that are not included in the Train set.
\end{itemize}

\subsection{Experiment Settings}
Following \cite{7}, we selected the same features they used, which are shown in table II. We would like to clarify that the Ports classification was performed manually, where ports were classified into 9 categories that were used in \cite{7}. It was done manually because it is dependent on the network and system settings more than number ranges.

\begin{table}[h]
\caption{Features Selected from NSL-KDD Data Set}
\begin{center}
\begin{tabular}{| p{2cm} | p{5.5cm} | p{2cm} | p{1cm} |}
\hline Feature & Description & Data Type & No. of Bins \\
\hline duration & Connection duration & Integer & 8 \\
\hline protocol\_type & Protocol type & Categorial & N/A \\ 
\hline service & Port category & Integer & 9 \\ 
\hline land & Land packet & Binary & N/A \\ 
\hline urgent & No. of urgent packets & Integer & 1 \\ 
\hline host & No. of "hot" indicators & Integer & 3 \\ 
\hline num\_failed\_logins & No. of failed login attempts & Integer & 3 \\ 
\hline logged\_in & If user logged in & Binary & N/A \\ 
\hline root\_shell & If root shell obtained & Binary & N/A \\ 
\hline su\_attempted & If command "su root" attempted & Binary & N/A \\ 
\hline num\_file\_creations & No. of file creations & Integer & 4 \\ 
\hline num\_shells & No. of open shell prompts & Integer & 2 \\ 
\hline is\_host\_login & If user in "hot list" & Binary & N/A \\ 
\hline is\_guest\_login & If user logged as guest & Binary & N/A \\ 
\hline count & No. of connections to same host in past 2 s & Integer & 10 \\ 
\hline same\_srv\_rate & \% of connections to same port in past 2 s & Real & 3 \\ 
\hline diff\_srv\_rate & \% of connections to different ports in past 2 s & Real & 3 \\ 
\hline srv\_diff\_host\_rate & \% of connections to different hosts in past 2 s & Real & 3 \\ 
\hline 
\end{tabular} 
\end{center}
\end{table}

The values used for the GA parameters were:
\begin{itemize}
\item 	Population size: 200, 400, 600
\item	Number of generations: 200, 500, 1000, 2000
\item	Mutation rate: 2/L, where L is the number of features.
\item	Crossover rate: 1.0
\end{itemize}

Different values of population size and number of generations were used to compare the results to see which would lead to better results.

\subsection{Results}
After running the algorithm on the train set normal samples, we had sets of detectors (rules) dependent on population size and number of generations. Running those detectors on the test set, the detection rates are shown in figure 1.2

\begin{figure}[h]
\begin{center}
{\includegraphics[width=4in,height=2.5in]{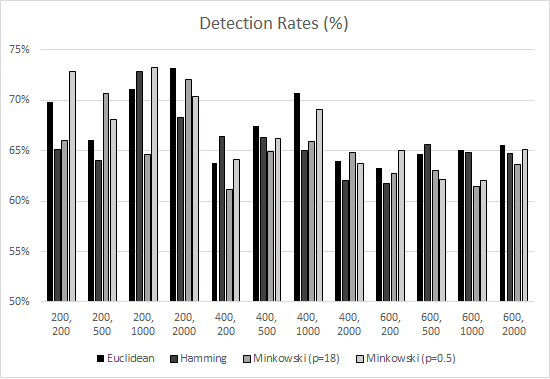}}
\caption{Test Set Detection Rates}
\end{center}
\end{figure}

As shown in figure 1.2, the detection rates are not very high - as the maximum detection rates realized are 73.23\% and 73.31\% obtained by the detectors generated by GA applying Euclidean and Minkowski (p=0.5) distances respectively. The rates with population size 200 are generally better. To measure the IDS efficiency, true positives and true negatives rates (TPR and TNR respectively) are calculated and shown in figures 1.3 and 1.4.

\begin{figure}[h]
\begin{center}
{\includegraphics[width=4in,height=2.5in]{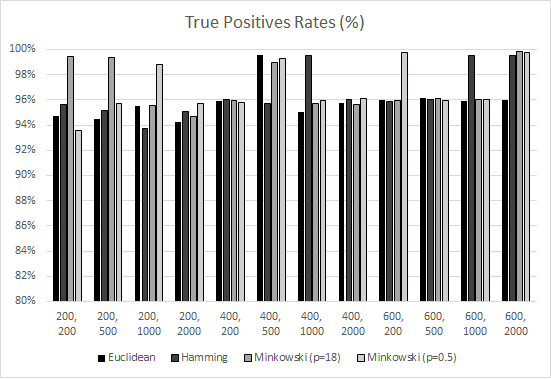}}
\caption{True Positives Rates}
\end{center}
\end{figure}

\begin{figure}[h]
\begin{center}
{\includegraphics[width=4in,height=2.5in]{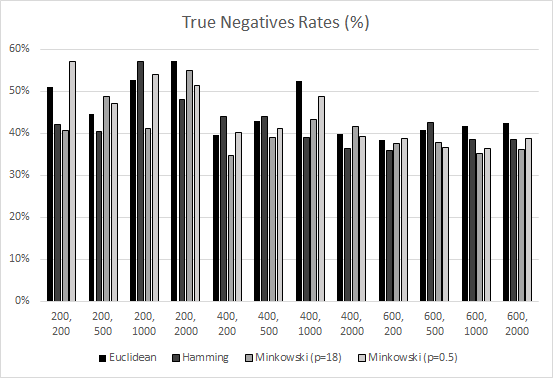}}
\caption{True Negatives Rates}
\end{center}
\end{figure}

We can realize in figure 1.3 that the recognition of normal samples is high as the TPRs are close to 96\% with FPRs all less than 7\%. The problem is the discrimination between normal and anomalous samples, as the TNRs (Figure 1.4) are all below 60\% with FNRs all above 40\%. So, the IDS is not able to detect all anomalies as some are seen as normal activities. Better TNRs are obtained using detectors generated using bigger population size, mostly with the Minkowski distance.

\section{Conclusion and Future Work}
In the paper, an algorithm is implemented to generate detectors that should be able to detect anomalous activities in the network. The data was pre-processed before using them in the algorithm, by discretizing the continuous features to create homogeneity between data values, by replacing values with bin numbers. Seeing the results, we can see that the equal-width interval binning algorithm that we used is very simple but not very efficient. As mentioned before, it's static and global, hence it doesn't take into consideration the relations between features and whether the predefined number of bins is efficient or not. As for the parameters of the GA, the detectors generated by GA with smaller population size gave better detection rates, true alarms, and lower false alarms that others generated using higher population sizes.Our future work will be focusing on applying other discretization algorithms that are more dynamic, and consider using other features to detect more anomalies.

\end{document}